\documentclass[preprint,12pt]{elsarticle}

\usepackage{amssymb}

\journal{JBIS (SI: 8th Interstellar Symposium)}

\begin{document}

\begin{frontmatter}

\title{Constraints on Interstellar Sovereignty}

\author[inst1]{Jacob Haqq-Misra}
\ead{jacob@bmsis.org}

\affiliation[inst1]{organization={Blue Marble Space Institute of Science},
            addressline={600 1st Avenue, 1st Floor}, 
            city={Seattle},
            state={WA},
            postcode={98104}, 
            country={USA}}

\begin{abstract}
Human space exploration and settlement of other planets is becoming increasingly technologically feasible, while mission concepts for remote and crewed missions to nearby star systems continue to be developed. But the long-term success of space settlement also requires extensions and advances in models of governance. This paper provides a synthesis of the physical factors that will constrain the application of sovereignty in space as well as legal precedent on Earth that likely applies to any crewed or uncrewed missions to other stellar systems. The Outer Space Treaty limits the territorial expansion of states into space, but the requirements for oversight of nongovernmental agencies and retention of property ownership enable the extension of state jurisdiction into space. Pragmatic constraints from historical precedent on Earth suggest that new space treaties will be unlikely to succeed and new global space agencies may have limited jurisdiction over states, while hard constraints of the space environment require adherence to technical capabilities, political feasibility, and long-term sustainability. These factors form a three-prong test for assessing the viability of interstellar governance models. This discussion of interstellar governance is intended to further the conversation about sovereignty in space prior to the first intentional launch of any interstellar spacecraft. 
\end{abstract}



\end{frontmatter}

\newpage


\section{Introduction}
\label{sec:introduction}

Growing interest among state and private space agencies in exploring interplanetary space has expanded the number of stakeholders engaged in robotic solar system missions as well as those intending to pursue crewed exploration of the moon or Mars. The stated goal of SpaceX founder Elon Musk to make humans and life ``multi-planetary'' \cite{musk2017making,musk2018making} reflects the long-term vision of a broader set of state and private stakeholders who are investing in the requisite technology. This momentum toward an expanded space economy and permanent human presence in space has also reignited discussions regarding eventual interstellar exploration. As the technical capability to conduct interstellar missions increases, parallel developments in management and governance will also be required.

One example is the Interstellar Probe mission concept \cite{mcnutt2022interstellar}, which would send a spacecraft to a distance of fifty astronomical units to study the outer heliosphere of the sun as well as the local interstellar medium. Another example is the Breakthrough Starshot concept \cite{parkin2018breakthrough}, which would use a laser to accelerate hundreds to thousands of tiny probes toward the Alpha Centauri system. Even more speculative are mission concepts for intergenerational worldships \cite[see, e.g.,][]{haqq2022future}, which would carry small human populations through interstellar space at subrelativistic speeds and sustain multiple generational cycles before reaching the destination. Due to the long travel times to reach interstellar destinations, such missions will require effective intergenerational management in order to sustain operations and achieve their intended objectives. But the existence of such mission concepts demonstrates the present interest in thinking about the possible objectives of interstellar missions, the technological feasibility of developing such missions in the near-term future, and the need for developing pragmatic governance models for interstellar exploration.

This paper provides an overview and synthesis of the legal and political factors that would constrain interstellar sovereignty. This discussion is based on limits that are provided by intertnational law, international relations, and pragmatism; the goal is to survey the existing legal and political environment that applies to cislunar and interplanetary space to understand the potential implications for interstellar missions. The structure of this paper on interstellar sovereignty draws upon the analysis of interplanetary sovereignty by \citet{haqq2022sovereign} in \textit{Sovereign Mars}, with the intention of highlighting areas in which interstellar exploration would face novel or unexplored governance challenges. The paper begins with discussion of legal constraints imposed by the Outer Space Treaty (Section \ref{sec:OST}), followed by a set of pragmatic constraints from historical precedent (Section \ref{sec:pragmatic}), and a set of hard constraints from the physical demands of the space environment (Section \ref{sec:hard}). These three considerations form a three-prong test for assessing the viability of potential governance models for interstellar exploration, as the long-term viability of interstellar missions will require a pragmatic approach that acknowledges the political and technical realities of space.

\section{Outer Space Treaty}\label{sec:OST}

The Outer Space Treaty (OST) of 1967, formally known as the Treaty on Principles Governing the Activities of States in the Exploration and Use of Outer Space, including the Moon and Other Celestial Bodies, provides the primary formal legal constraints on sovereignty in space. The OST has 115 parties today, which includes all states with spacefaring capabilities. Twenty-two states have signed but not ratified the treaty, with the greatest nonparticipation among Sub-Saharan African states. Attempts at extending the provisions of the OST include the Moon Agreement of 1979, formally known as the Agreement Governing the Activities of States on the Moon and Other Celestial Bodies, which has today has seventeen states as parties. Although the Moon Agreement remains binding on these parties, none of these are spacefaring states with launch capabilities, so this is considered by most scholars to be a failed treaty \cite{hertzfeld2015simple}. The OST thus is the primary multilateral international treaty that applies to sovereignty in cislunar, interplanetary, and interstellar space.

\subsection{Expansion}

The OST limits the territorial expansion of sovereign states into space and declares space as free for exploration and use by all states. These provisions are described in Articles I and II, which apply equally to all regions of space. The OST does not provide any unique guidance for interstellar space, as interstellar travel was a remote possibility when the treaty was drafted, but instead the wording of Articles I and II applies to any region of space, which should apply equally to interplanetary as well as interstellar space.

Article I of the OST requires that ``the exploration and use of outer space, including the Moon and other celestial bodies'' is the ``province of all mankind,'' which requires ``free access'' on a ``basis of equality'' and includes ``freedom of scientific investigation.'' The phrase ``province of all mankind'' requires that the activities of exploring and using of space must be ``for the benefit and in the interests of all countries,'' so that no state can deny another access to conduct such activities in space; nevertheless, legal scholars generally do not interpret Article I as an obligation for technology transfer or economic redistribution from spacefaring to nonspacefaring states. One of the reasons for the failure of the Moon Agreement was the attempt to classify the moon and celestial bodies in the solar system (other than Earth) as the ``common heritage of mankind'' and obligated economic redistribution from spacefaring activities under governance of a new international regime. Common heritage principles remain absent in the OST, so Article I maintains the equality of access to space for all without any further obligations between states.

 The OST limits the extent to which sovereign states on Earth can claim territory in space. The text of Article II states that space ``is not subject to national appropriation by claim of sovereignty, by means of use or occupation, or by any other means.'' Article II prevents any state on Earth from extending its national boundaries to include objects or resources in space. The OST was drafted during the heightened geopolitical tensions of the Cold War, and the language of Article II reflects in part the concerns at the time regarding the possible expansion of Soviet or American forces into space. The OST does not necessarily prohibit the act of landing on planetary bodies or establishing long-term settlements, but Article II states that any act of settlement, resource use, or sustained occupation does not qualify as a claim to ownership or sovereignty. Likewise the act of discovery, first arrival, or other method of territorial claims cannot justify exclusive use to any planetary bodies in space. Article II does not necessarily prevent scientific activities---such as sample return---or economic use of space resources---such as asteroid mining---and so interstellar exploration missions would likewise be able to utilize local resources as needed; however, economic use of space resources still must be distinguished from claims to sovereignty over territory in space.

\subsection{Weaponization}

Limits on the weaponizing of space are described in Article IV of the OST, which includes different provisions for cislunar orbit compared to interplanetary or interstellar space. For objects in orbit around Earth, Article IV prohibits ``nuclear weapons or any other kinds of weapons of mass destruction'' but does not necessarily exclude other types of weapons. Many astronauts have carried knives to space for utility purposes, while cosmonauts landing in Siberia were issued pistols to protect themselves from wildlife as they awaited rescue. Various efforts to develop defensive weapons in space, such as the Strategic Defense Initiative of 1983, do not necessarily violate Article IV unless they would qualify as a weapon of mass destruction. 

 Beyond Earth orbit, including the moon and other planets, Article IV requires space to be used ``exclusively for peaceful purposes,'' with ``the testing of any type of weapons'' as well as ``military manoeuvres'' being ``forbidden'' by all states. This article makes exceptions for military personnel engaged in peaceful scientific research and also states that ``any equipment or facility necessary for peaceful exploration'' is also permitted. Scant legal precedent exists that could clarify the extent to which defensive weapons could qualify as ``necessary'' for peaceful exploration in space. Article IV prohibits ``military bases, installations, and fortifications'' on celestial bodies but also permits the use of ``military personnel'' for scientific or peaceful purposes. Interplanetary and interstellar missions could include significant military participation, and might even include the transport (but not testing) of weapons, but any such activities could not lead to the establishment of a permanent military outpost on another planet. 

\subsection{Oversight}

The OST also holds states responsible for the activities of their own private space agencies and are likewise liable for any associated damages to other states. These provisions are described in Articles VI, VII, and VIII, which apply equally to all regions of space. States are also given the right to inspect the space installations of other states under Article XII. Although Article II prohibits a formal expansion of state sovereignty into space, the provisions in Articles VI, VII, VIII, and XII all provide mechanisms for extending state jurisdiction through space exploration and settlement.

 Article VI gives states the responsibility for activities in space, which applies equally to ``activities carried on by governmental agencies or by non-governmental agencies,'' for the purpose of ensuring compliance with the OST. State governments are further obligated to provide ``authorization and continuing supervision'' to non-governmental entities that operate in space. States with launch capabilities currently provide authorization for any space missions developed or launched from within their sovereign territory, but the requirement for ``continuing supervision'' remains unclear when pertaining to long-term space exploration. Many commercial space agencies are beginning to develop capabilities that rival some state space agencies, so it remains conceivable that some states may be limited in the extent to which such supervision could occur; likewise, long-term interplanetary missions, such as Mars settlement, may not be practical for government agencies to supervise through continuous monitoring \cite{haqq2022sovereign}. Interstellar missions further complicate the requirement for continuing supervision, as private entities engaged in interstellar space exploration may be utilizing novel technology that few state agencies hold. Remote supervision can occur to an extent, but the first near-term efforts at interplanetary and interstellar exploration will begin to establish practices for how such continuing supervision will be conducted.

 States that authorize the launch of objects into space from their sovereign territory are liable for any damage to another state or its ``natural or juridical persons'' caused by such objects or their components, according to Article VII of the OST. This includes liability due to damage to another state that may occur to another state at the time of launch, damage that occurs while in space, and any damage associated with atmospheric re-entry and landing. This liability can extend to both the state in which a launch occurs as well as the state in which an private operator is incorporated, which applies even if one state is a rogue actor. It is also worth noting that the Article VII does not address domestic liability from objects launched into space. Further provisions governing liability between states for objects launched into space were codified by the Liability Convention of 1972. Interstellar missions face the same liability risks at the time of launch, but interstellar missions are most likely to be one-way expeditions that do not typically return a crew or payload back to Earth. The potential liability for damage to another state is also significantly lower for interstellar missions when compared to objects in Earth orbit.

 Situations could arise in which a private space agency is incorporated in one nation but launches from another. In such scenarios, the state that hosts and authorizes the launch would be liable for any damages associated with launch, flight, and re-entry, according to Article VII. At the same time, the state in which a private space agency is incorporated would be responsible for authorizing and overseeing the company’s activities in space, according to Article VI. This could lead to legal complications in which property damages in space could be considered either a liability issue, falling upon the launch state as per Article VII, or an oversight issue, falling upon the state of incorporation as per Article VI. A remote possibility also exists that a private space agency could choose to launch from one of the few states that has not ratified the OST, known as the rogue state scenario; however, such a scenario would likely have negative political consequences for the actors involved.

 States retain jurisdiction over property and personnel that are launched into space, according to Article VIII. Although Article II prohibits the sovereign appropriation of territory in space, Article VIII states that ``ownership of objects launched in to outer space, including objects landed or constructed on a celestial body, and of their component parts, is not affected by their presence in outer space.'' Likewise, the state of registry ``shall retain jurisdiction and control over such object[s], and over any personnel thereof, while in outer space or on a celestial body,'' according to Article VIII. Further details regarding the registry of objects launched into space were prescribed by the Registration Convention of 1976. This enables states to extend their influence beyond Earth and into space through space missions and space settlement; even though this cannot constitute an extension of territorial sovereignty, such actions still extend the jurisdiction of a state beyond Earth. The governance challenges in implementing Article VIII are already evident in missions such as the International Space Station, which contains modules owned by five different nations—in which each nation’s laws uniquely apply. As an example, states participating in the International Space Station maintain a series of bilateral and multilateral agreements and memorandums of understanding in order to manage the shared technical infrastructure and flow of digital information across the space station \cite{sadeh2004technical}. Multinational interstellar missions would likewise face additional complications under Article VIII to ensure a legal regime on the mission that reflects the association between state jurisdiction and property ownership.

 The OST also requires states to make space installations, stations, equipment, and vehicles open to inspection by representatives of other states ``on a basis of reciprocity,'' according to Article XII. Any state representative seeking to make such an inspection visit under Article XII must provide ``reasonable advance notice of a projected visit'' in order to ``assure safety'' and ``avoid interference with normal operations in the facility.'' In principle, Article XII prevents a state from maintaining exclusive access to a space station or other installation, regardless of the purpose of construction; however, such scenarios have not yet occurred, so the legal conditions under which a state can engage in such an inspection remain unknown. The practice of reciprocal inspection visits may become more relevant as China and private space agencies develop space stations, as well as the longer-term vision of state and private agencies to establish permanent settlements on Mars. The language of Article XII refers to ``visits'' rather than remote monitoring, but it may be infeasible for states to inspect the interstellar equipment of other states by sending a representative. Interstellar missions may face fewer demands for reciprocal inspections, particularly if states choose to prioritize different interstellar targets for exploration.

\subsection{Disclosure}

The OST prioritizes the use of space for peaceful purposes and cooperative scientific exploration, which includes some obligations on the part of states to share information. These provisions are described in Articles V and XI, which apply equally to all regions of space. Cooperation and sharing of information is encouraged in most cases but required in certain circumstances.

Article V of the OST designates astronauts as ``envoys of mankind'' and requires states to ``render all possible assistance'' to the astronauts of other states in the event of an emergency in space, on planetary bodies, or landing on Earth. Article V also requires states to inform other states and the United Nations Secretary-General of any discoveries in space that ``could constitute a danger to the life or health of astronauts.'' The extent to which a state could conceal a known risk in space may depend on the state's assessment of the likelihood of other astronauts being exposed to the particular risk factors. Interstellar missions will face similar issues regarding the duty to disclose dangers discovered in interstellar space, which may not threaten the astronauts of other states on any near-term timescales. Likewise, the language of Article V provides protection specifically to ``astronauts'' in space, which may or may not include the broader class of space tourists that pay to visit space but are not actively contributing to a space mission \cite{lyall2010astronaut,langston2015name}. Precedent for such law will likely emerge as space tourism industries develop a larger base of customers, so such ambiguities regarding the extent of protection provided by Article V to passengers aboard space missions may be resolved prior to any crewed interstellar missions.

Article XI of the OST includes an agreement by states to share the ``nature, conduct, locations, and results'' of activities in space. The text of Article XI promotes such disclosure of activities and information ``to the greatest extent feasible and practical''. This stipulation conceivably provides an option for a state to withhold or restrict information from space activities if disclosure would reveal state secrets, create security threats, or pose other risks to the state. The extent to which a state could be held liable for failing to disclose potentially harmful space activities remains untested in international law. Interstellar missions may face some practical challenges to the rapid dissemination of information due to the communication delay between interstellar missions and Earth; nevertheless, such circumstances would not necessarily prevent the timely dissemination of information, unless additional steps were taken to withhold or securitize the information after sufficient internal analysis has occurred.

 The disclosure requirements under Articles V and XI can typically be fulfilled through the dissemination of scientific results in publications and conferences or through an international political body such as the UN Committee on the Peaceful Uses of Outer Space (COPUOS). However, these requirements also suggest situations in which a state could make a significant discovery in space but withhold any information from other states. For example, the discovery of extraterrestrial technology would be of broad interest to all states and all people, with the potential to bring benefits or cause harm \cite{michaud2007contact,baum2011would}, but the requirement of states under the OST to share such evidence could depend on the specific scenario of discovery \cite{Long2020}. Under Article V, disclosing the discovery of extraterrestrial technology would only be required if it would pose a danger to astronauts in space; even if ``astronaut'' is taken to apply to all humans, Article V would not apply to the discovery of extraterrestrial technology that is determined to be harmless by the discovering state. Under Article XI, such a discovery could be kept secret if announcing it would be infeasible or impractical; a state may determine that withholding the discovery of extraterrestrial technology would jeopardize national security or create mass panic, for example. Interstellar exploration will increase the ability to search exoplanetary systems for biosignatures and technosignatures that could indicate present or past life, and the duty to disclose such information under the OST may depend on the circumstances of the discovery and the actors involved.

\subsection{Preservation}

The OST also addresses potential contamination that could occur during space exploration. Article IX of the OST requires states to avoid both ``harmful contamination'' of space and planetary bodies---known as forward contamination---and ``adverse changes'' on Earth from returned material---known as backward contamination. The possibility of forward contamination is of greatest relevance to interstellar missions, which are likely to be one-way journeys and unlikely to include sample return.

Article IX of the OST does not provide further guidance regarding the extent of changes that would qualify as harmful or adverse, but the Committee on Space Research (COSPAR) has developed a set of planetary protection standards in consultation with international stakeholders that remain consistent with Article IX. COSPAR planetary protection policy \cite{coustenis2023planetary} specifies bioburden limits for missions that depend on the type of mission (flyby, orbiter, or lander) and the target planetary body, with stricter requirements for missions that might disrupt any extant life or regions that could host native life. Most state space agencies adhere to COSPAR planetary protection policies to comply with Article IX, although COSPAR standards are non-binding. States are also arguably responsible for ensuring compliance for their private space agencies, under Article VI. Current COSPAR policy provides detailed sterility requirements for robotic exploration of the solar system as well as general guidelines for minimizing contamination in the human exploration of Mars. These guidelines for crewed exploration are relatively recent additions to COSPAR planetary protection policy, and such guidelines may be further expanded and refined as the launch a near-term crewed Mars mission becomes more likely.

Current COSPAR planetary protection policies have no specific requirements for interstellar missions. Current spacecraft on interstellar trajectories---such as the Voyager, Pioneer, and New Horizons spacecraft---all were intended to explore solar system objects. The Interstellar Probe mission concept \cite{mcnutt2022interstellar} is an example of a mission that is intended to enter interstellar space, and if the mission does not conduct any planetary flybys during its voyage, then such a mission would likely not require any specific planetary protection measures under current COSPAR policy. Similarly, the Breakthrough Starshot initiative \cite{parkin2018breakthrough} would only need to consider planetary protection measures if its proposed flyby exploration of the Alpha Centauri system were considered a possible risk to environments that could host chemical evolution or life. Changes to COSPAR planetary protection policies occur regularly in consultation with space agencies and other stakeholders, and such policies may respond accordingly as interstellar missions become more likely. Such policies may technically still fall under the umbrella category of ``planetary protection,'' but eventually it may be useful to distinguish between policies intended for interplanetary protection of solar system planets and those intended for interstellar protection of exoplanets---especially if interstellar destinations are ultimately intended for long-term settlement.

Other suggestions have been made for extending efforts to preserve the space environment beyond the focus of current COSPAR policies on protecting regions of space likely to harbor or develop life. For example, \citet{cockell2004planetary,cockell2006planetary} have suggested that the recognition of a ``planetary parks'' system could designate certain areas of the martian surface, or other bounded regions in space, as ``wilderness'' to remain undisturbed. The designation of certain areas of space as planetary parks could provide a way to balance the interests of commercial, preservationist, and other stakeholders \cite{profitiliotis2023future}. The idea of planetary parks could apply directly to interstellar exploration, with regions of particular exoplanets designated as wilderness to be preserved. The planetary parks concept could also be extended into the concept of ``interstellar parks'' where entire exoplanets or even full exoplanetary systems are designated as wilderness regions with limited to no in situ exploration permitted. Recognizing planetary parks at interplanetary and interstellar scales would uphold the requirements of Article IX by limiting other forms of contamination or interference of the space environment beyond COSPAR policies.

Aside from the OST, additional environmental protections of the atmosphere and space were codified in the the Environmental Modification Convention of 1978, formally the Convention on the Prohibition of Military or Any Other Hostile Use of Environmental Modification Techniques. The Environmental Modification Convention has 78 parties today, which includes major spacefaring states, and prohibits ``military or any other hostile use of environmental modification techniques having widespread, long-lasting or severe effects as the means of destruction, damage or injury to any other State Party'' (Article I). This treaty defines  ``environmental modification techniques'' as ``any technique for changing—through the deliberate manipulation of natural processes—the dynamics, composition or structure of the Earth, including its biota, lithosphere, hydrosphere and atmosphere, or of outer space'' (Article II). The applicability of the Environmental Modification Convention to problems in space is largely untested: some scholars have noted that the growing accumulation of orbital space debris can be a source of damage by one state to another, but the extent to which the Environmental Modification Convention will be invoked in space debris cases, or other instances of pollution in space, remains to be seen \cite{steele2021space}.

Ethical considerations can also provide insight on applying preservationist ideas to space. The tradition of environmental ethics has discussed the possibility that non-human organisms or even non-living entities could hold intrinsic value, which can include zoocentric ethics (with animals holding intrinsic value) and biocentric ethics (with the biospehre holding intrinsic value). One particular approach for extending this valuation model to space is planetocentric ethics \cite{sullivan2013}, which suggests that planets and other objects in space are examples of ``nature's projects'' \cite{hargrove1984beyond} that could be preserved in proportion to their uniqueness. A planetocentric ethic would not necessarily prohibit space exploration but may suggest limiting large-scale transformation of planets that uniquely highlight physical properties unobserved elsewhere. Interstellar exploration may find greater relevance in planetocentrism as exoplanetary systems are explored through in situ methods, which may provide insight into the extent to which stellar systems host planets with unique physical characteristics. 

\section{Pragmatic Constraints}\label{sec:pragmatic}

No other strong legal constraints exist on interplanetary or interstellar sovereignty beyond the OST, but some potential limitations based on historical precedent are worth considering. The analysis by \citet{haqq2022sovereign} in \textit{Sovereign Mars} examined models of cooperative sovereignty on Earth, such as the UN Law of the Sea Convention and the Antarctic Treaty System, to identify a set of pragmatic constraints on cooperative sovereignty in interplanetary space. These three pragmatic constraints are:
\begin{quote}
(1) the emergence of new international organizations with jurisdiction over space activities will face opposition from some or all of the major spacefaring states;\\
(2) any requirements for mandatory equitable sharing of space resources will be unable to gain complete participation by all spacefaring states; and\\
(3) any new space treaties will be unable to attract sufficient signatories among spacefaring states to become relevant as international law. \cite[][215--216]{haqq2022sovereign}    
\end{quote}
This set of pragmatic constraints represents a working hypothesis for the conditions that are likely to surround the emergence of new space law. But these constraints may not necessarily be correct, and it remains possibly that any or all of these conditions could be significantly different in the future. These constraints should therefore be considered as a starting place for thinking about pragmatic limits to interstellar exploration, while acknowledging that future conditions at the time of such missions could be significantly different. The text below provides a brief overview of the basis for these three constraints, bur readers are encouraged to examine the full exposition and supporting references in \textit{Sovereign Mars} \cite[][chapter 6]{haqq2022sovereign} for a more thorough discussion.

The first constraint limits the emergence of a world space agency, interplanetary authority, or interstellar federation that would hold jurisdiction over existing sovereign states. Although such models of unilateral governance across interplanetary and interstellar distances may be prevalent in science fiction, historical precedent on Earth suggests a resistance of states to cede their sovereignty to new external authorities. As one example, the present version of the Law of the Sea Convention has failed to gain global participation---with notable non-participants including Israel, Turkey, the United States, and Venezuela---due to the establishment of the International Seabed Authority as a new juridical body to govern the use of seafloor resources. The objection of these states arises in part because of the reluctance to grant jurisdiction to the International Seabed Authority that would otherwise reside at the national level. As another example, the Antarctic Treaty System was developed outside of the jurisdiction of the UN, with participation initially limited to states engaged in Antarctic exploration at the time, and participation currently remains open to any state that engages in Antarctic science. Delegates from the UN initially expressed concerns regarding the ``Question of Antarctica,'' although such issues were gradually resolved as the Antarctic Treaty invited broader participation. Nevertheless, the Antarctic Treaty System remains an independent framework that is governed by the consultative parties to the treaty, without granting further authority to existing international governance organizations like the UN. This constraint implies that interplanetary exploration will hold national sovereignty as the ultimate source of authority. This does not necessarily mean that international organizations will be ineffective but only that any international organizations or agreements that develop will likely not require a state to yield its sovereignty to a new interstellar authority. 

The second constraint notes that historical attempts at mandating equitable sharing or other required redistribution of resources have failed to gain complete participation by all states. Spacefaring states in particular may be unlikely to support mandates for equitable sharing, an objection that arises in part because of the investment burden borne by the first states to develop the technology for spaceflight. As one example, the failure of the Moon Agreement occurred largely because its Article 11 designated the moon and its resources as ``the common heritage of mankind,'' which would be governed by a new ``international regime''. The purpose of establishing the international regime under Article 11 includes ``an equitable sharing by all States Parties in the benefits derived from those resources''. Whereas the OST succeeded in establishing free access for the exploration and use of space by states, the Moon Agreement failed to require equitable sharing of space resources. As another example, the objections of some states to joining the Law of the Sea Convention arose because of the requirements for equitable sharing of resources extracted from the deep seabed, which the treaty describes using common heritage principles. Concerns over required technology transfer by the International Seabed Authority are one example of an obstacle that would be faced by attempts at requiring equitable technology transfer for interstellar exploration. The lack of mandatory redistribution does not preclude voluntary redistribution or collaboration, but this constraint does suggest that interstellar missions will be carried out for the benefit of national taxpayers, private investors, or other specific stakeholders, but without strong obligations for global redistribution.

The third constraint limits the significance of new space treaties as sources of international law, as non-participation by some of the major spacefaring states would render such treaties ineffective. The Moon Agreement is one example of an attempted treaty that has still not been ratified by any states with launch capabilities, while no other significant space treaties have yet emerged within the UN framework. Meanwhile, national policies in the United States, Luxembourg, Japan, and the United Arab Emirates have authorized the extraction of space resources through mining, which asserts the claim of such states that these space activities would not constitute ``national appropriation'' under Article II of the OST. Multilateral agreements such as the Artemis Accords remain consistent with this constraint, as such agreements are not necessarily legally binding and do not make strong obligations of participants. Other commercially-driven and grassroots approaches toward space governance may emerge through common practice, which could ultimately lead to the recognition of such practices as customary international law. Much of maritime law developed as the recognition of existing practices, and later codification thereof, so similar practices could occur in space without the need for new international  treaties. Interstellar exploration will similarly not receive any further guidance from UN multilateral treaties, which gives a larger role to voluntary partnerships and unilateral agreements among states, international nongovernmental organizations, groups of commercial stakeholders, and other public forums for developing models for interstellar governance.

\section{Hard Constraints}\label{sec:hard}

The legal and pragmatic constraints discussed so far are soft limits that could conceivably differ in the future, but physical limitations based on the requirements for survival in space provide some hard constraints on interstellar sovereignty. The analysis by \citet{haqq2022sovereign} in \textit{Sovereign Mars} described three hard limits on sovereignty in interplanetary space. These three hard constraints are (emphasis added):
\begin{quote}
Any model for governance on Earth that is attempted on Mars should demonstrate the \textit{technical capability} to develop and manage any required infrastructure in order to be considered viable. Likewise, any such model should demonstrate \textit{political feasibility} given the interests and momentum of existing stakeholders. Finally, any terrestrial governance model used on Mars would need to ensure \textit{long-term sustainability} by developing effective management practices that keep consumption within the system's carrying capacity. \cite[][218]{haqq2022sovereign}    
\end{quote}
This set of hard constraints cannot be ignored when attempting to develop pragmatic governance models for interplanetary or interstellar space travel. Hard constraints acknowledge that advances in technology and changes in geopolitics will inevitably lead to a future in which certain options may be possible regarding technology, politics, or sustainability that are not presently available. Hard constraints do not eliminate ideas or technologies that are plausible; instead, hard constraints eliminate scenarios that are impossible as well as those that have no plausible continuity from the present. Such hard constraints will impose even greater challenges for interstellar missions that must effectively manage operations across generations in order to succeed.

Technical capability represents a hard constraint for any infrastructure, and the challenges of surviving in the space environment intensify the requirements for long-term space missions. Commodities on Earth such as breathable air must be manufactured in space, with associated costs that may limit the extent to which permanent settlements could grow \cite{stevens2015price}. Long-term missions must also develop methods for utilizing local space resources; interplanetary missions could rely upon supplemental provisions from Earth deliveries (which for a martian settlement could occur about every twenty-six months when Earth and Mars are closest), but interstellar missions would need the technical capabilities to remain self-sufficient. The infrastructure required for life support must also be resilient to a range or risks to ensure that critical systems can maintain operation without interruption. Sustained support will also be required to ensure the longevity of any long-term space missions. Interstellar missions in particular will have limited contact with Earth, with the time delay between transmission and receipt increasing with time as the spacecraft's distance increases. Maintaining and managing large-scale projects across intergenerational timescales also remains challenging for human civilization, and interstellar missions will require unprecedented strategies for intergenerational succession of management, transfer of knowledge expertise, and longevity of technical infrastructure.

Political feasibility requires that any model for governance in space must acknowledge the present political reality, which includes the interests and momentum of existing stakeholders. The acknowledgment of political feasibility does not neglect the possibility of change, even radical change in the future, nor does it require that the interests of existing stakeholders must necessarily dictate future trajectories. Instead, the hard constraint of political feasibility requires that any governance models that would require major political shifts to succeed must include a plausible trajectory between the present and intended future. In other words, any future scenarios for governance cannot be ungrounded to the extent that the required future could not in any way be derived from present conditions. This approach would rule out many of the ``soft'' science fiction scenarios for space governance as well as idealistic models that may not be politically feasible under any realistic circumstances. The requirement for political feasibility also considers the historical context in which space exploration emerged, which includes a modern continuation of national competition that began during the Cold War, juxtaposed with the interests of private space agencies in developing a profitable space economy. Scientific justifications for space missions in many ways provide a vehicle for extending national sovereignty into space, while remaining consistent with the OST, so scenarios in which scientific exploration is completely decoupled from political interests may be untenable. Interstellar governance will be further complicated by communication delays between sending and receiving a transmission, due to the light-year scale separation between Earth and any interstellar destinations. Such challenges may inevitably require some  degree of autonomy for decision-making on interstellar missions.

Long-term sustainability represents a concern for any human endeavor, but the demands of the space environment and dependence on built infrastructure pose even greater challenges for the success of interplanetary space settlement or interstellar travel. The ecological concept of carrying capacity describes the maximum population size that an environment can sustain, which in space depends entirely on life support technology. The physical area required to house a population in space will be limited by the size of the settlements or spacecraft, while food production in space can only support a population of a finite size. Such considerations will drive hard limits on populations in space, which will require a net zero growth rate that balances births, deaths, and other factors that would affect a space population \cite{tiwari2021factors,milligan2015rawlsian}. Immigration and emigration could also be factors for interplanetary settlements, such as the periodic arrival of migrants from Earth to Mars, but immigration would not be relevant for interstellar travel. The engineering concept of a safety factor describes the extent to which a system can exceed its intended load, which for space missions includes the ability to support more personnel than present on the crew. Such safety factors enable mitigation against risk, and utilization of the excess capacity usually reflects a problem in the mission. Physical constraints on infrastructure and resources demand a strategy for long-term sustainable management in order for efforts at permanent interplanetary settlement or interstellar exploration to succeed. Interplanetary exploration in particular, even uncrewed missions, will face numerous sustainability challenges to ensure longevity of the technology and availability of sufficient energy for maintaining operations across intergenerational timescales.

\section{Conclusion}

The idea of interstellar exploration raises numerous questions regarding sovereignty and governance in space that extend and heighten similar concerns for interplanetary exploration. The limits on interstellar exploration imposed by the OST, pragmatic constraints from history, and hard physical constraints form a three-prong test for assessing the viability of interstellar governance models. This three-prong test does not necessarily require that a given model account for all constraints, but instead the test provides a way for comparing various governance models with one another and assessing their plausibility for future implementation. This test applies to both crewed and uncrewed interstellar missions, as both mission types involve the extension of state jurisdiction by sending property (if not personnel) into space.
	
Compliance with the OST is not necessarily mandatory for a plausible governance model. Once a state is bound by a treaty, it has three options for how to proceed: adhere to the treaty, withdraw from the treaty, or ignore the treaty. Noncompliance with the requirements of a treaty can create enforcement challenges, particularly for states that refuse to acknowledge the jurisdiction of international court systems. Although ignoring the OST may not be a prudent political strategy today, future interstellar missions may operate in a different political environment in which the interpretation or recognition of the OST has changed.

The pragmatic constraints discussed in Section \ref{sec:pragmatic} serve as a working hypothesis based on successful and failed examples of cooperative sovereignty from history, but this working hypothesis could be incorrect. Historical precedent does not necessarily imply likelihoods about future developments, and it remains possible that new space treaties could emerge or that spacefaring states could support the emergence of new sovereign political entities in space. But such speculations cannot be boundless, and additional sets of pragmatic constraints may be useful to identify from historical, philosophical, or sociological analyses that could guide the development of interstellar governance models.

The hard limits discussed in Section \ref{sec:hard} cannot be violated, to the extent that such limits represent physical impositions of the space environment. Some breakthrough technologies could conceivably alter hard limits, but such breakthroughs cannot necessarily be assumed to be typical, nor can they be anticipated to solve critical problems of management or infrastructure. The physical realities of the space environment, including complete dependence on life support technology for crewed missions, insist upon pragmatic governance models to ensure safety and continuity of the mission. 

Future governance models for interstellar exploration must include some degree of pragmatism. The idea of interstellar exploration---and speculation about interstellar settlement---can stimulate the mind to imagine limitless possibilities for such future scenarios. This vast set of imagined possibilities is rich territory for thought experiments in governance, but only a subset of such scenarios will be relevant for interstellar mission planning.

\section*{Acknowledgments}
Thanks to Upasana Dasgupta and an anonymous reviewer for comments that significantly improved this paper. The author gratefully acknowledges support from the NASA Exobiology program under grant 80NSSC22K1009. Any opinions, findings, and conclusions or recommendations expressed in this material are those of the author and do not necessarily reflect the views of any employer or NASA.

\bibliographystyle{elsarticle-num-names} 
\bibliography{refs}

\end{document}